# PRELIMINARY STEPS IN DESIGNING AND IMPLEMENTING A PRIVILEGE VERIFIER FOR PMI

Diana Berbecaru, Antonio Lioy
*Politecnico di Torino,*
*Dip. di Automatica e Informatica,*
*Torino, ITALY*

**ABSTRACT**

We have designed and deployed a system that uses X.509 public-key certificates (PKC) and attribute certificates (AC) for access control. This includes an authorization service for on-line environments where clients are identified by X.509 PKCs and their privileges are expressed with X.509 ACs. During a request to a protected resource, a privilege verifier decides if the user satisfies all the requirements to get access to the controlled resource. In this paper we focus on the steps to be performed by the privilege verifier, which is the entity in charge of validating both the PKCs and the ACs involved. The validation of PKCs and of ACs are two separate tasks but they are closely related. In our system we have identified two distinct entities: the privilege verifier whose task is to validate ACs, and the certificate validation server, whose task is to validate the PKCs. The validation of PKC is more complex, and it thus can be implemented and provided as a service by a dedicated authority, named Validation Authority. This paper describes the model, architecture and implementation of this system. It also includes some preliminary measurements and our future plans for the development of the system.

**KEYWORDS**

PMI, attribute certificate, privilege verifier, certificate validation.

## 1. INTRODUCTION

The 2003 annual "Computer Crime and Security Survey" [1] states that most organizations employ some measure of access control (92%) but still 56% of the respondents reported an unauthorized use of computer systems within the last 12 months and a $70,195,900 loss from the theft of proprietary information. An important technology that can be used for access control is the Privilege Management Infrastructure (PMI) [2]. The main function of a PMI is to provide a strong authorization control after the authentication phase. The research and development efforts done in the PMI area are still in the beginning phase.

The ACs, issued by Attribute Authorities, are derived from PKCs (or identity certificates) and were introduced to respond to the requirement to offer a robust, distributed and scalable system for the management of authorizations. In fact, the PKCs can contain attribute information to be used for authorization purposes. The main drawback related to the insertion of attributes (e.g. the role of a person in an organization) in the PKCs is that the roles change more often than the duration of a PKC and this leads to the necessity to revoke and issue again the certificate. Thus, as the PKCs proved to be inadequate to store privilege information, the international standards community responsible for the PKC format [2] developed a similar certificate format without the public key information and for the express purpose of storing privilege information, named ACs. Like PKCs, however, ACs require an infrastructure to manage the certificates throughout their lifecycle. This infrastructure is commonly referred to as a PMI. An authorization service can be designed using ACs which each point to a PKC.

The X.509 standard specifies that, in a PMI, a *privilege asserter* must present an AC containing the appropriate attributes/privileges to a *privilege verifier* before access is granted to an information object. The *privilege verifier* acts as a reference monitor and controls access to the object. The Internet AC Profile for authorization [3] provides the algorithm the privilege verifier must execute for AC validation. The decision to allow access is based also on the security policy being enforced by the verifier and any applicable environment variables (e.g. time of day).

Several architectures for distributed access control that are based entirely on certificates were proposed, such as Akenti [4, 5] or Permis [6]. These are rather complex systems that describe the management of different types of certificates for access control. Nevertheless we have not found in the literature a system that details the drawbacks and functionality of a privilege verifier as described in [3]. Thus, when X.509 digital certificates and/or attribute certificates are used for authorization we considered that further work has to be employed for the privilege verifier subsystem. This is motivated also by the observation that a privilege verifier can be used either in on-line or off-line environments. For example, in an on-line web server the privilege verifier is usually implemented as a module that is integrated with the web server to control the client browser's access toward the web server's resources. On the other hand the privilege verifier can be embedded at the user part itself and thus be a component of an off-line environment. For example it can be part of a document viewing software, such as a PDF viewer. Some solutions have been proposed for on-line environments, such as the Akenti policy engine that can be integrated with the Apache web server to process all necessary certificates. An off-line environment is typically controlled and configured by the user. Therefore, all necessary data must be downloaded and configured in advance on the user side. In [7] the authors explain a scenario where, besides the root attribute authority that issues ACs to customer entities, the PMI consists also of a PMI-enabled PDF viewer. Embedded in the PDF viewer is the public key of the root attribute authority. The PDF viewer uses this public key to verify the authenticity of a user's attribute certificate. The user's AC contains his or her right-to-execute for a given title of a document distributed by a company that wished to sell documents online. The PDF viewer also has a copy of a root certificate authority's public key to verify the identity of the user. We can notice thus that in the viewer are embedded at least the root keys for the certificate authority and the attribute authority. In other document viewing systems, additional information could be required, such as the certificate revocation lists (CRLs). In this case the privilege verifier designer must carefully identify all necessary configuration parameters and the management of certificate revocation update/checking.

In this paper we propose a privilege verifier system for on-line environments that makes use of X.509 digital certificates and attribute certificate for access control. The main role of the privilege verifier is to validate all necessary data to restrict the access to the resources. Specifically, in our system we focus on the validation both of the identity and of the attribute certificates defined by the X.509 standard and the profile proposed by the IETF PKIX working group for the above two types of certificates [3, 8]. The algorithm described in [3] for AC validation states that when a PKC is to be used it must be validated too. Attribute certificate validation and public-key certificate validation are different tasks but they are closely correlated since the AC validation employs PKC validation. Our contribution is the design and implementation of a *privilege verifier* in on-line environment that uses a dedicated service for X.509 PKC validation. In our system this task is executed by a server, named certificate validation server, and is not present in the access control systems presented above. The privilege verifier proposed behaves as an application gateway between the certificate-enabled (web) clients and the validation server. The paper is structured in the following way: Section 2 points out the related work, Section 3 describes a study model used to identify the privilege verification operations, Section 4 details our proposed privilege verification subsystem and its functionality, Section 5 describes the prototype implementation and Section 6 gives the conclusions and the future work.

## 2. RELATED WORK

### 2.1 Solutions and Open Issues in Certificate-Based Access Control

Permis [6] developed a role-based access control information infrastructure that tries to solve identification and authorization problems and uses X.509 ACs to store the user's role. Permis privilege verification subsystem distinguishes between the application-specific Access Control Enforcement Function (AEF) and the application-independent Access Control Decision Function (ADF). The AEF authenticates the user, and then asks the ADF if the user is allowed to perform the requested action on the particular target resource. ADF bases its decision on information retrieved from one or more LDAP directories to retrieve the authorization policy and the role ACs for the user. In Akenti [4], the heart of the system is the Akenti policy engine, which gathers and verifies certificates and then evaluates the user's right to access the requested resource based on these certificates. The model includes a resource server that interfaces to resources on behalf of the client, the policy engine that must verify the so-called use-condition certificates that apply to a resource, the caching server that caches all types of certificates used in the system (i.e. use-condition certificates, identity certificates and attribute certificates) and the log server.

Even if privileges could be delegated from one entity to another by means of ACs as defined in [2], the resulting delegation chains proved to be hard to handle. Authors investigated in [9] how the ITU-T standard X.509 can be used to support the Canadian Department of National Defence authority and delegation models. The results provide insight into and quantification of the complexity of the resulting delegation chains. The conclusions of the authors are that public-key operations are expensive and the complexity of implementing their model seems high. This confirms the complexity warnings expressed in [3] where Farrell and Housley do not recommend the use of delegation chains.

To mitigate the complexity for retrieving and processing the certificate paths, a cache could be used within the privilege verifier components. But this solution is not appropriate for reasons explained in [10]. In this paper authors argue that the benefits of using both identity and access control certificates have an unanticipated consequence in practice, namely, that of having to support *selective* and *transitive* certificate-revocation and access review. With selective revocation the paper states that a dependency must be enforced between the use of attribute (group) certificates and the identity certificate during certificate revocation. Thus, if the identity certificate is revoked (and thus included in the CRL) but the attribute certificate is not revoked (not present in the ACRL) a principal may be able to retain access in an unauthorized manner and later repudiate it. Consequently, selective revocation requires that the AC of a principal whose identity certificate has been revoked (and was used to obtain the attribute certificate) be selectively revoked as well. This case is particularly relevant in applications where principals are registered in different infrastructure domains serving different user coalitions, or *ad-hoc* networks with overlapping and dynamically varying membership. Another issue raised in the paper is the cache invalidation, i.e. there must exist a method to immediately invalidate the cache locations where are stored the PKCs and/or ACs. The authors explain how an intruder can get access to a restricted object because there is always an effective delay between the revocation of the identity certificate and the access revocation. One solution proposed to this problem is to validate all concerned identity certificates (PKCs) at every request. We'll explain briefly in the next section the steps to be performed in this case.

## 2.2 Mechanism and Open Issues in X.509 Public-Key Certificate Validation

We'll review briefly here the steps that need to be performed in order to validate a PKC. One of the first steps is *parsing* and *syntax checking* of the digital certificate and its contents, including some semantic checking like use of certificate compared to allowed use and presence of mandatory fields and critical extensions. Another step is the *validation of the CA's signature* on the PKC. This requires a trusted copy of the CA's own public key. If the CA is not directly trusted a certificate chain must be constructed up to a trusted anchor: this process is called certificate path construction. The certificate path construction process uses the *Subject* and *Issuer* PKC fields and the matching of the values of the *Authority Key Identifier* and *Subject Key Identifier* extensions [11]. When a certification path is built from the end entity to a trusted anchor, this is called building in the forward direction. When a certification path is built from a trust anchor to the end entity, this is called building in the reverse direction [12]. According to the PKI trust model, the best method for building certification path is different. For example, for hierarchical or trust list models, the best way is to build the certification path in the forward direction. Another step is a *check* that the PKC is within its validity period, indicated by the *not before* and *not after* PKC fields. Another important step is to *check* that the certificate is not *revoked*, i.e. declared invalid by the CA before the end of its validity period. Revocation may appear because the certificate holder is no longer entitled to have the certificate, or because of misuse, compromise or loss of the private key. A further step is the *semantic* processing of the certificate content. This means that information is extracted from the certificate and is further presented to the relying party (through a user interface or as parameters) for further processing by the application. This should include an indication of the quality of the certificate and the CA, based on the identification of the certificate policy that the CA applied for certificate issuance. The above processing is described as four steps in [8]: (1)

Initialization, (2) Basic certificate checking, (3) Preparation for the next certificate in the sequence, (4) Wrap-up. Steps (1) and (4) are each accomplished once, step 2 is performed for every certificate in the chain and step 3 is performed for all certificates in the path except the last certificate in the chain, the end-entity certificate. Currently, there exist several libraries that implement totally or partly the PKC validation, such as Certificate Management Library, OpenSSL and Microsoft Crypto API.

The PKIX working group defined the requirements [13] and examined various candidate validation protocols such as SCVP [14], DVCS [15] and OCSP [16], that can be used in client-server architectures where the client can delegate partially or totally the path processing to a server in order to minimize the client complexity. The server can return a variety of information, such as certificate status information, certification path and the validity of the PKC. Initially thought to be used mostly by thin clients, the server proved to be useful also for centralized management of trust relationships and certificate policies. However the above protocols are only protocols that support certificate validation. They don't specify mechanisms such as retrieving certificate status (they assume that the revocation information is already available in some way, either from a CRL or from a database back-end), building certification path, managing certificate policy, and so on. Moreover, for some of the protocols (such as DVCS, OCSP) further extensions must be defined in order to support PKC validation [17, 18].

## 3. PRIVILEGE VERIFIER OPERATIONS

The aim of our privilege verifier system is to support security critical applications with valid information about the current status of the users' access rights (privileges). To identify the steps that the privilege verifier must perform to validate an AC according to the algorithm in [3], we'll use the simplified model in Figure 1. We don't take into consideration at the moment more complex scenarios, such as when the client has more than one $PKC_{Client}$ or $AC_{Client}$ issued by different authorities or that CA1 and CA2 are part of a more complex PKI model.

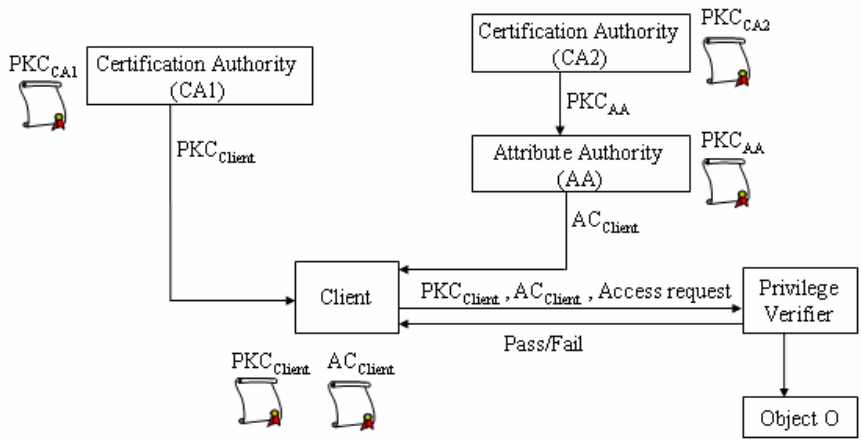

Figure 1. Simplified model used for analyzing privilege verifier operations

The privilege of a privilege holder reflects the degree of trust placed in that holder, by the certificate issuer, that the privilege holder will adhere to those aspects of policy that are not enforced by technical means. The privilege is encapsulated either in the privilege holder's AC(s) or in the *subjectDirectoryAttributes* extension of its PKC, which may be presented to the privilege verifier in the invocation request. The privilege verifier's task is to determine that the privileges of the holder are valid and sufficient to invoke a specific method of an object. The pass/fail decision taken by the privilege verifier depends on the following factors: privilege of the privilege asserter, privilege policy, environment variables and some parameters of the invoked method. When the privilege is encapsulated as *subjectDirectoryAttributes* extension of the client's PKC, named $PKC_{Client}$, the task of the privilege verifier is to validate $PKC_{Client}$. Alternatively, the PKCs and the ACs can be used together in order to execute role-based authorization: the user identifies himself using the $PKC_{Client}$ and then the system uses further the $AC_{Client}$ in the authorization process. The preliminary phase of authentication is fundamental since it must avoid that the privileges contained in the $AC_{Client}$ be associated to a false end entity. The algorithm described in [3] for AC validation states that when a PKC is to be used it must be validated too. Thus, with respect to the PKI, the privilege verifier must validate $PKC_{Client}$ according to the procedure described shortly in Section 2.2. If the verifier does not directly trust $PKC_{Client}$ it must construct and validate the certificate chain ($PKC_{Client}$, $PKC_{CA1}$). Next, the privilege verifier has to validate AA's digital signature on $AC_{Client}$. For this purpose, the referenced public key $PKC_{AA}$ must be checked for its validity. We can note thus that the following two types of information need to be configured on the privilege verifier to allow PKC validation: trust anchors and revocation data. In the model the AA has its certificate $PKC_{AA}$ issued by a certification authority, which we called CA2 in the Figure 1, and the verifier needs the means to be able to accept CA2. Thus, the $PKC_{CA2}$ must be configured on the privilege verifier as a trusted anchor so that the privilege verifier can construct and validate the certificate chain ($PKC_{AA}$, $PKC_{CA2}$). There is no necessary connection between the CA that issued $PKC_{Client}$ (CA1 in our model) and the CA that certified the AA (CA2 in our model). Consequently, this will require the privilege verifier to fetch and construct two certificate chains ($PKC_{Client}$, $PKC_{CA1}$) and ($PKC_{AA}$, $PKC_{CA2}$) to validate the credential represented by $AC_{Client}$. Consequently, $PKC_{CA1}$ also has to be configured as trust anchor on the privilege verifier. One important step of the PKC validation is revocation checking. The standard methods for revocation checking are CRLs [8] and OCSP [16] and the privilege verifier must check the PKC status with one of the above methods. We can observe that the complexity of the privilege verifier increases if it has to support other PKI models.

## 4. PROPOSED PRIVILEGE VERIFIER SYSTEM

Figure 2 illustrates the proposed privilege verifier subsystem. We can distinguish basically two entities: the privilege verifier and the certificate validation server (CVS). We'll note further that the operations the privilege verifier performs are simpler while the complexity of the CVS is higher. When the privilege verifier system is used in an organization, we can assume that it is managed by the organization while the CVS is managed by an external entity (named Validation Authority) that provides the certificate validation service to the organization.

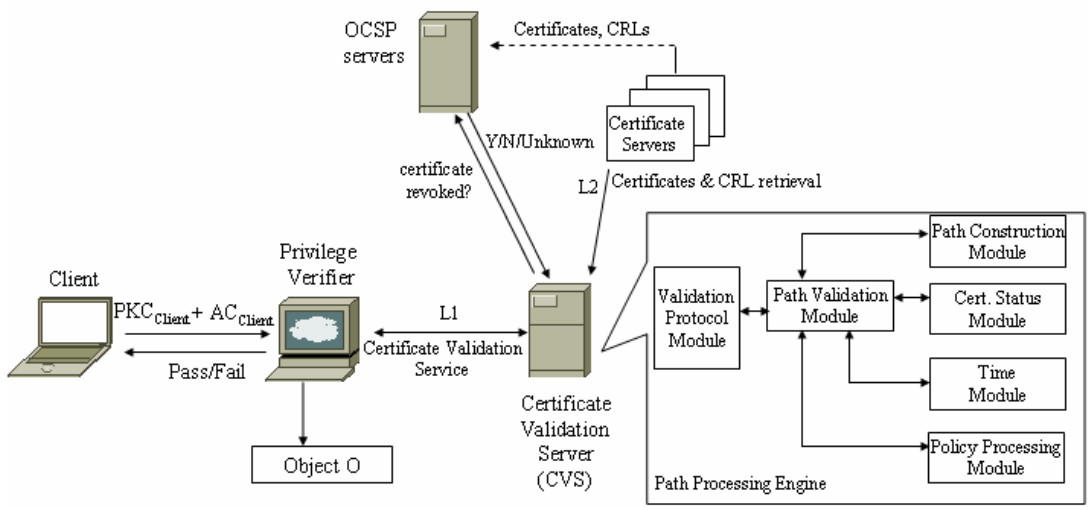

Figure 2. Privilege verifier subsystem components

Since in our implementation the privileges are expressed by ACs, the privilege verifier has basically the task to validate the ACs and for this purpose it performs the following operations:
- control the *formal* validity of the AC. This step is referred to in [2] as basic processing procedure.
- control the validity relative to the service. At this step the privileges corresponding to the attributes contained in the AC are verified.

In our system the delegation chains are not supported. The control of formal validity consists in establishing if the certificate respects a number of conditions, independently of the entity that performs the

321

verification and that can impose further restrictions or particular requirements. At this step the privilege verifier will check that:
- the AC has not been altered. This means that the signature applied on the AC must be correct and the public-key certification paths related to the AC must be correctly validated. In our system, the privilege verifier communicates with the certificate validation server to process the public-key certification paths. For this purpose it embeds a module acting as a client in the communication with CVS.
- the AC was not revoked (in our proposal, the privilege verifier uses CRLs stored locally).
- the time of evaluation is in the period of validity of the AC.
- whether or not the privileges being asserted are sufficient for the context of use, according to the rules established by the privilege policy that specifies the environment variables to be considered.

The control of validity relative to the service is performed after the above steps of the control of formal validity have completed. Each *privilege verifier* can adopt a different policy to distribute its services. It is thus necessary to verify first that the verifier recognizes the attribute extensions marked as critical. After that it evaluates the extensions (like AC Targeting) that can limit the validity of AC only to a restricted set of services. In the third place it must further check that in the AC are present the attributes that the service provider has established as necessary in order to allow access to its services.

The CVS runs a *Path Processing Engine*, which is not viewed as an ad-hoc application, but was further split in several components, each one having a dedicated task. Thus, the *Path Validation Module (PVM)* executes the path validation algorithm to determine the validation status of the target PKC. The algorithms for path validation are included in this module. The role of the *Path Construction Module (PCM)* is to construct certification paths. The algorithms for path development, which use certificate extensions and loop detection/elimination techniques to increase efficiency, are embedded here. The *Certificate Status Module (CSM)* determines the status of a certificate. Thus, the code for retrieving and processing CRLs is contained in *CSM*. If OCSP is supported, an OCSP client is integrated here to communicate with the responders. The *Policy Processing Module (PPM)* treats policies and can be further divided into submodules: one to process the certificate policies and another to process the validation policies. The *Time Module (TM)* provides an accurate and reliable time reference. The *Storage Module* is used to store/retrieve certificates, certificate revocation data and policies. The Communication Module (*CM*) handles the communication part, e.g. sockets and the SSL channels. The typical server application on CVS receives and parses the requests by making calls to the Validation Protocol Module (*VPM)*. The values contained in the request fields are passed as input to the *PVM* or *PCM*, which will make further calls to functions from the other modules. The *Path Processing Engine* is not based on a particular crypto or certificate management library to allow developers to freely choose the underlying support, as well as the support for DER encoding/decoding of structures handled. In our system the flow of the PKC/AC validation service is as follows:

1) The client tries to access the protected object and for this purpose it sends $PKC_{Client}$ and $AC_{Client}$ toward the privilege verifier.
2) The privilege verifier executes verification operations on the $AC_{Client}$ as explained above. When it has to validate one PKC, the privilege verifier passes it to the module that constructs the certificate validation request, i.e. the VPM module. The format of the request depends on the protocol chosen for the link L1. The certificate validation request may optionally include the client's trust anchor or the certification policy and in some cases, it can contain also the certificates used for constructing the certification path. This information is used for validating the PKC. Optionally, the privilege verifier could be configure to digitally sign the overall request.
5) The constructed request is passed to the *CM* module from where it can be sent on unsecure or on an SSL channel to the *CM* module on CVS.
6) The server builds PKC certification paths using the *PCM* module. If the trust anchor is present in the request, the server must build the certification path that starts from the trust anchor certificate.
7) If building certification path succeeded in step 6) and the certificate policy is present, the server verifies certification path using certification policy constraints and also performs checking of the certificate status.
8) If building certification path succeeded in step 6) and the certification policy is not present, the validation process is performed by only checking the certificate status with the *CSM* module.
9) An optional step here would be to store in a cache on the CVS either the certification path or the certificate status information (CRL, OCSP responses) so that they would be available at the next processing.

## 5. PROTOTYPE AUTHORIZATION SERVICE

The testbed used for the implementation of a prototype authorization service based on our system is composed of a PC Pentium III 700 MHz running RedHat 7.2 with SSL-enabled Apache version 1.3.20 for the web server and privilege verifier. The CVS runs on a PC Pentium III 1 GHz with Windows 2000 Professional and uses the OpenSSL library to support SSL communication channels between the privilege verifier and the CVS and for managing the PKC certificates and the requests/response exchanged on L1. The Apache server is configured to support client and server authentication and to export the client and server certificates as environment variables. Although there are no special requirements about the client (e.g. any browser supporting cookies can be used) we used Mozilla 1.5.

In the current implementation, a verification script in Perl, named *verify.pl*, is invoked by the client and thus the verification script acts as the interface for the application that validates the ACs (named *ACVerify*) and the module that constructs the certificate validation request to be sent to CVS (named *PKCVerify*). The verification script is invoked via a SSL connection with client authentication and the client browser must have $AC_{Client}$ stored in some way (e.g. as a cookie). The script is called when the user tries to access a protected resource, such as an HTML page named *resource.htm*. When the user tries to connect with https to *resource.htm* the client browser sends $PKC_{Client}$ during SSL connection set up and it sends the $AC_{Client}$ as a cookie. If the execution of the verification script runs with success, the server loads the required page, otherwise an error message is returned. Since the web server is configured to require client authentication, at the end of the SSL handshake the web server has an authenticated X.509 $PKC_{Client}$ for the user. The value of $PKC_{Client}$ is obtained from the environment variable SSL_CLIENT_CERT and is passed in as parameter to the *PKCVerify* that constructs the request and sends it further to the CVS to be validated. The data required for the execution of the verification script, is obtained from three sources:

- the $AC_{Client}$, stored as a cookie.
- information on $PKC_{Client}$ returned by the web server. This information, such as the validity date of the $PKC_{Client}$, the values of the fields holder or issuer and serial number, will be used also for the verification of the $AC_{Client.}$
- variables hard-coded in the verification script or read from a configuration file. The script or the configuration file contains a list of variables that must be set on the privilege verifier. These variables contain for example the path towards *ACVerify* and *PKCVerify*, the policy file, the list of trusted authorities, which attributes must be verified to consider valid the $AC_{Client.}$

The *PKCVerify* run by the privilege verifier is a standalone client application whose task is to construct the certificate validation request and send it to the CVS. On the CVS, the server receives the certificate validation requests through the VPM module, then extract the necessary data and pass it to the module in charge of PKC validation. Examples of data extracted from the request are the PKC to be validated, the trust anchor indicated by the privilege verifier and other intermediary certificates. The protocol used for L1 is the DVCS protocol [15] and for validation of PKCs we used the Certificate Management Library [19].

For testing purpose the $PKC_{Client}$ was issued by Politecnico CA (http://ca.polito.it). Additionally we implemented an Attribute Authority that issues the $AC_{Client}$. When CVS is not used for PKC validation it took on average 0.003 s on the web server to distribute a 1K page and 1.70 s for a 1 Mbyte page on a SSL channel between the client and the privilege verifier. When CVS is used through an SSL channel on link L1 it took on average 1.17 s to distribute a 1K page. From this time, it took about 0.4 s for CVS to construct and validate the path of length 4 for $PKC_{Client}$, and 0.04 s to verify the $AC_{Client}$. The numbers are illustrated in Table 1.

Table 1. Average times to fetch a document from the privilege verifier system

| File size | No CVS  | CVS and L1 with SSL |
|-----------|---------|---------------------|
| 1 KB      | 0.003 s | 1.17 s              |
| 1 MB      | 1.700 s | 2.90 s              |

In the tests we measured only the time for the execution of PKC validation function *CM_RetrieveKey* [19] consequently the additional load necessary for CML library initialization was not taken into account. Also we used the options for searching locally the certificates and CRLs in the two files *cert.db* and *CRL.db* that are created by default at CML library initialization and used throughout the PKC validation. The rest of 0.7 seconds was used to construct DVCS requests and responses and the communication time in a 100 Mb/s LAN. For digitally signing the DVCS requests/responses we used RSA digital signatures. On average we measured that it takes 0.01 s for constructing and 0.03 s for digitally signing a DVCS request of 2356 bytes and 0.04 s for constructing a DVCS response of 1403 bytes.

## 6. CONCLUSIONS AND FUTURE WORK

To control the access to a resource, both authentication and authorization are needed. Early versions of the ITU-T X.509 standard have concentrated on standardizing strong authentication techniques, based on PKCs. The latest version of X.509, published in 2002, is the first edition to standardize an authorization technique based on ACs and PMIs. In this paper we presented a privilege verifier system that uses PKCs and ACs for authorization in on-line environments and we detailed its functionality. Moreover, since the *privilege verifier*, the entity in charge with the validation of ACs, must also verify the identity of every entity in the path using the certification path procedure identified in the X.509 standard, this will increase considerably its complexity. For this purpose, in our proposal, the *privilege verifier* uses an external service for the validation of PKCs. Future work includes optimization of the two main components of the system: on the privilege verifier our intent is to identify the requirements specific to off-line environments and adapt the current implementation to these requirements, on the CVS we intend to integrate an algorithm for revocation checking based on OCSP with fallback on CRL and that makes use of a local cache on CVS.

## REFERENCES

[1] CSI/FBI Computer Crime and Security Survey, http://www.gocsi.com, 2003.

[2] ITU-T Recommendation X.509: *Information Technology - Open Systems Interconnection - The Directory: Public Key and AttributeCertificate Frameworks*, March 2000.

[3] S. Farrell, R. Housley, *An Internet Attribute Certificate Profile for Authorizations*, IETF, RFC-3281, 2002.

[4] W. Johnston, S. Mudumbai, M. Thompson, *Authorization and attribute certificates for widely distributed access control*, Proc. of 7th IEEE Workshop on Enabling Technologies: Infrastructure for Collaborative Enterprises, pp: 340-345, 1998.

[5] M. Thompson, W. Johnston, S. Mudumbai, G. Hoo, K. Jackson, A. Essiari, *Certificate-based access control for widely distributed resources*, Proc. of 8th USENIX Security Symposium, pp: 215-228, 1999.

[6] D.W. Chadwick, O. Otenko, *The PERMIS X.509 Role Based Privilege Management Infrastructure*. Seventh ACM Symposium on Access Control Models and Technologies , pp: 135-140, 2002.

[7] T. Kohno, M. McGovern, *On the Global Content PMI: Improved Copy-Protected Internet Content Distribution*. Proc. of the Financial Cryptography, Fifth International Conference, 2001.

[8] R. Housley, W. Polk, W. Ford, D. Solo, *Internet X.509 Public Key Infrastructure Certificate and CRL Profile*, IETF, RFC-3280, 2002.

[9] S. Knight, C. Grandy, *Scalability Issues in PMI Delegation*, Proc. of the 1st Annual PKI Research Workshop, pp: 77-87, 2002.

[10] H. Khurana, V. Gligor, *Review and Revocation of Access Privileges Distributed with PKI Certificates*, vol. 2133 of *Lecture Notes in Computer Science*. Security Protocols, 8th International Workshop, Cambridge, 2000.

[11] S. Lloyd, *Understanding certification path construction*, whitepaper, PKI forum, 2002.

[12] Y. Elley, A. Anderson, S. Hanna, S. Mullan, R. Perlman, S. Proctor, *Building certification paths: Forward vs. reverse*, Proc. of Network and Distributed System Security Symposium, 2001.

[13] D. Pinkas, R. Housley, *Delegated Path Validation and Delegated Path Discovery Protocol Requirements*, IETF, RFC-3379, September 2002.

[14] A. Malpani, R. Housley, T. Freeman, *Simple Certificate Validation Protocol*, IETF, Internet draft, June 2002.

[15] C. Adams, P. Sylvester, M. Zolotarev, R. Zuccherato, *Data Validation and Certification Server Protocols*, IETF, RFC-3029, 2001.

[16] M. Myers, R. Ankney, A. Malpani, S. Galperin, C. Adams, *X.509 Internet Public Key Infrastructure Online Certificate Status Protocol (OCSP)*, IETF, RFC-2560, 1999.

[17] GPKI Application Implementation Guide Report, *Development of Interoperable Assistive Technology for Digital Government Information Security*, available at http://www.jnsa.org/mpki/.

[18] D. Berbecaru, A. Lioy, *Towards Simplifying PKI Implementation: Client-Server based Validation of Public Key Certificates*, Proc. of IEEE International Symposium on Signal Processing and Information Technology, pp: 277-282, 2002.

[19] Certificate Management Library, http://digitalnet.com/knowledge/cml_home.htm

# PROCEEDINGS OF THE
# IADIS INTERNATIONAL CONFERENCE
# WWW/INTERNET 2004

Volume I

MADRID, SPAIN

OCTOBER 6-9, 2004

Organised by
**IADIS**
**International Association for Development of the Information Society**